\documentstyle[ltwol]{article}

\def \beq{\begin{equation}}
\def \eeq{\end{equation}}
\def \bea{\begin{eqnarray}}
\def \eea{\end{eqnarray}}
\def \s{\sqrt{2}}

\def\fbar{\overline{f}}

\def\tl{\tan\theta_c}

\def\to{\rightarrow}

\begin{document}
\title{RECENT THEORETICAL DEVELOPMENTS IN CP VIOLATION}
\author{MICHAEL GRONAU}
\address{Department of Physics, Technion - Israel Institute of Technology, 
32000 Haifa, Israel}

\twocolumn[\maketitle\abstracts{We review recent suggestions for testing 
through $B$ decays the flavor structure of CP violation in the Standard Model.
Relative signs of CP asymmetries in U-spin related processes can by themselves
test the Kobayashi-Maskawa mechanism in a crude manner. Ratios of 
charge-averaged decay rates and certain CP asymmetries may  
constrain tightly the weak phases $\gamma=\phi_3$ and $\alpha=\phi_2$.}]

\section{Introduction}
Assuming that a phase in the Cabibbo-Kobayashi-Maskawa (CKM) matrix is
the {\it dominant} source of CP violation, sizable CP asymmetries are predicted
in a variety of $B$ decay processes. A major task of present
experiments at $e^+e^-~B$ factories, following an earlier measurement
by the CDF Collaboration at the Fermilab Tevatron \cite{CDF}, is the search for
a time-dependent asymmetry in $B^0(t)\to J/\psi K_S$, which can be cleanly 
interpreted in terms of the weak phase $\beta=\phi_1$ \cite{BS,MG}. As a 
consequence of measurements reported at this conference by the BaBar 
\cite{BaBar} and Belle \cite{Belle} Collaborations, the present world average 
value of $\sin 2\beta$ gained the respectable status of a nonzero measurement 
at three standard deviation, $\sin 2\beta = 0.48 \pm 0.16$, consistent with the 
CKM prediction. The remarkable success in running the two asymmetric $B$ 
factories at SLAC and KEK-B, to be soon joined by experiments at the Tevatron, 
promises a considerable improvement in this important measurement during the 
coming year. 

In addition to time-dependent asymmetries, such as in $B^0(t)\to J/\psi K_S$ or 
$B^0(t)\to \pi^+\pi^-$, which are related to CKM phases, $B$ decays provide an 
opportunity of measuring CP rate asymmetries in a large number of two body and 
mutibody $B$ decay modes. In this talk we choose to discuss two classes 
of decay processes. In the first class, discussed in Section 2, asymmetries can 
provide crude but very useful tests of the CKM mechanism. While the signs and 
magnitudes of such direct asymmetries depend in general on unknown strong final 
state phases, these asymmetries will be shown to be related to each other 
pairwise in some approximation. The relative signs of pairs of asymmetries in 
this large class of processes are predicted quite reliably, implying that 
{\it crude} asymmetry measurements can provide simple tests. Measuring ``wrong" 
relative signs would most likely imply new physics. 

In a second class of processes studied in Section 3 (which in some cases 
overlaps with the first class) {\it precise} asymmetry measurements are shown 
to provide tight constraints on CKM phases. In several cases such information 
may be gained by merely measuring certain ratios of charge-averaged decay rates. 
Combining all these methods allows for precision tests of the CKM hypothesis of 
CP violation. Such tests will hopefully provide first clues for physics beyond 
the Standard Model. 

Our brief review \footnote{Invited talk at BCP4, Ise-Shima, Japan, February 
19--23, 2001} will focus on recent work, discussing
several central examples which represent a much broader study made during the 
past decade \cite{B2k}.
 
\section{A theorem about equal CP rate differences}
A subgroup of flavor SU(3), discrete U-spin symmetry interchanging $d$ and $s$ 
quarks, plays a particularly powerful and important role in charmless $B$ 
decays~\cite{Uspin}. Consider the low energy effective weak Hamiltonian 
describing $\Delta S =1$ charmless $B$ decays~\cite{BBL}:
\bea\label{Heff}
{\cal H}^{(s)}_{\rm eff} &=& \frac{G_F}{\s}\left [ V^*_{ub}V_{us}
\left ( \sum^2_1 c_i Q^{us}_i +\sum^{10}_3 c_i Q^s_i\right ) \right.
\nonumber\\
&+& \left. V^*_{cb}V_{cs}\left(\sum^2_1 c_i 
Q^{cs}_i +\sum^{10}_3 c_i Q^s_i\right )\right]~,
\eea
where $c_i$ are scale-dependent Wilson coefficients. The flavor structure
of the various four-quark operators is $Q^{qs}_{1,2}\sim\bar b 
q\bar q s,~Q^s_{3,..,6}\sim \bar b s \sum \bar q' q',~Q^s_{7,..,10}\sim 
\bar b s\sum e_{q'}\bar q' q'.~e_{q'}$ are quark charges,~$q'=u,d,s,c$.
Each of the four-quark operators represents an $s$ 
component (``down") of a U-spin doublet, so that one can write in short
\beq\label{Us}
{\cal H}^{(s)}_{\rm eff} = V^*_{ub}V_{us}U^s + V^*_{cb}V_{cs}C^s~,
\eeq
where $U$ and $C$ are U-spin doublet operators. Similarly, the effective 
Hamiltonian responsible for $\Delta S =0$ decays, in which one replaces 
$s\rightarrow d$, involves $d$ components
(``up" in U-spin) of corresponding operators multiplying CKM factors
$V^*_{ub}V_{ud}$ and $V^*_{cb}V_{cd}$,
\beq\label{Ud}
{\cal H}^{(d)}_{\rm eff} = V^*_{ub}V_{ud}U^d + V^*_{cb}V_{cd}C^d~.
\eeq

This structure of the Hamiltonian implies a general relation between
two decay processes, $\Delta S=1$ and  $\Delta S =0$, in which initial and 
final states are obtained from each other by a U-spin transformation,
$U: d\leftrightarrow s$. Writing the $\Delta S=1$ amplitude as
\beq\label{s}
A(B\to f,~\Delta S =1) = V^*_{ub}V_{us}A_u + V^*_{cb}V_{cs}A_c~,
\eeq
the corresponding $\Delta S =0$ amplitude is given by
\beq\label{d}
A(UB\to Uf,~\Delta S =0) = V^*_{ub}V_{ud}A_u + V^*_{cb}V_{cd}A_c~.
\eeq
Here $A_u$ and $A_c$ are complex amplitudes involving CP-conserving phases.
The amplitudes of the  corresponding charge-conjugate processes are
\beq\label{sbar}
A(\bar B\to \fbar,~\Delta S =-1) = V_{ub}V^*_{us}A_u + V_{cb}V^*_{cs}A_c~,
\eeq
and
\beq\label{dbar}
A(U\bar B\to U\fbar,~\Delta S =0) = V_{ub}V^*_{ud}A_u + V_{cb}V^*_{cd}A_c~.
\eeq
Unitarity of the CKM matrix \cite{Jarl}, ${\rm Im}(V^*_{ub}V_{us}V_{cb}V^*_{cs}) 
= - {\rm Im}(V^*_{ub}V_{ud}V_{cb}V^*_{cd})$,
implies the following relation between CP rate differences \cite{Uspin}
\bea\label{asym}
\Delta(B \to f) & \equiv & \Gamma(B \to f) - \Gamma(\bar B\to \fbar) \approx 
 \\
-\Delta(UB \to Uf) & \equiv & - [\Gamma(UB \to Uf) - \Gamma(U\bar B \to 
U\fbar)]~.\nonumber
\eea 

Namely, {\em CP rate differences ($\Delta$) in decays which go into one another 
under 
interchanging $s$ and $d$ quarks have equal magnitudes and opposite signs}.
This rather powerful result, following from U-spin within the CKM framework, 
can be demonstrated in numerous decay processes, including two body, quasi-two 
body, multibody hadronic and radiative $B$ decays. A few examples are
\bea
\Delta\Gamma(B^0\to K^+\pi^-) &\simeq& -\Delta\Gamma(B_s\to \pi^+ K^-)~,
\nonumber\\
\Delta\Gamma(B^0\to K^{*+}\pi^-) &\simeq& -\Delta\Gamma (B_s\to \rho^+ K^-)~,
\nonumber\\
\Delta\Gamma(B^+\to K^+\pi^+\pi^-) &\simeq& -\Delta\Gamma (B^+\to \pi^+ K^+ 
K^-)~,
\nonumber\\
\Delta\Gamma(B^+\to K^{*+}\gamma) &\simeq& -\Delta\Gamma (B^+\to \rho^+ 
\gamma)~.
\eea

U-spin is an approximate symmetry of strong interactions. Naively one would
think that the approximation (\ref{asym}) holds up to small terms of order 
$m_s/m_b$. Considering $B$ decays to two light pseudoscalars, and assuming 
\cite{BBNS,Li,Brod} that the dominant terms in amplitudes factorize 
\cite{Silves}, 
U-spin breaking corrections in these processes are given in terms of 
ratios of decay constants and form factors involving $s$ and $d $ quarks. 
This may lead to violations of asymmetry relations, however such violations are 
not expected to be gross. Furthermore, independently of any assumption, if such 
asymmetries are large, corresponding to large final state phases in a particular 
process, it is very unlikely that U-spin breaking can change the sign of these 
phases. Consequently, {\it the prediction that large CP asymmetries in two 
U-spin related proceesses have opposite signs is expected to be robust even in 
the presence of U-spin breaking effects}.

Can one correct (\ref{asym}) for U-spin breaking effects? Since such effects 
are model-dependent, it would be very useful if they could be directly measured  
in rates. There exists such a possibility \cite{Uspin} if one assumes that 
certain rescattering effects can be neglected. Consider the three pairs of U-spin
related processes ($B^0\to K^+K^-,~B_s\to\pi^+\pi^-$),~($B^0 \to K^+\pi^-,
~B_s \to \pi^+ K^-$) and ($B^0 \to \pi^+\pi^-,~B_s \to K^+ K^-$). The three
$\Delta S =1$ decays are described by the following SU(3) flavor flow 
amplitudes \cite{GHLR}: 
\bea\label{amps}
A(B_s \to\pi^+\pi^-) &=& - PA -E~,\nonumber \\
A(B^0 \to K^+\pi^-) &=& -P - T - \frac{2}{3}P^c_{EW}~,\\
A(B_s \to K^+ K^-) &=& -P - T -\frac{2}{3}P^c_{EW} - PA - E~\nonumber.
\eea
The corresponding strangeness conserving decay amplitudes involve other 
CKM factors but have the same SU(3) structure. 

The first pair of U-spin
processes involve only quark amplitudes $PA + E$ in which, in the absence 
of large rescattering effects \cite{rescat}, the spectator quark in the $B$ or 
$B_s$ meson participates in the interaction~\cite{GHLR}. These decays
can be used to test the smallness of rescattering corrections. Unless amplified 
by rescattering, these amplitudes are expected to be suppressed by $f_B/m_B$ 
relative to the dominant amplitudes occuring in the other two pairs of 
processes. Thus,
the branching ratio of $B^0\to K^+K^-$ is expected to be of the order of 
$10^{-7}$ or smaller, compared to $10^{-5}$ characterizing the branching ratios 
of the other four processes. {\it In order to test the assumption of small 
rescattering effects, the present experimental upper limit \cite{pipi}, 
${\cal B}(B^0\to K^+K^-) < 1.9\times 10^{-6}$, should be improved by at least 
one order of magnitude.}

Assuming that such a stringent bound is obtained, one can then neglect 
corresponding $PA + E$ terms in $B_s \to K^+ K^-$ and $B^0 \to \pi^+\pi^-$. 
This implies in the limit of U-spin symmetry 
\bea\label{SU3}
A(B_s \to K^+ K^-) & \simeq &  A(B^0 \to K^+\pi^-)~,
\nonumber\\ 
A(B_s \to \pi^+ K^-) & \simeq & A(B^0 \to \pi^+\pi^-)~.
\eea
The rates of these four processes can be used to measure U-spin 
corrections. For instance, assuming factorization these corrections are given
by ratios of form factors
\bea\label{su3}
\frac{A(B_s\to K^+ K^-)}{A(B^0\to K^+\pi^-)} &=& \frac{F_{B_s K}(m^2_K)}
{F_{B\pi}(m^2_K)}~,
\nonumber \\
\frac{A(B_s\to K^-\pi^+)}{A(B^0\to \pi^+\pi^-)} &=& \frac{F_{B_s K}(m^2_{\pi})}
{F_{B\pi}(m^2_{\pi})}~.
\eea
The two ratios of form factors are expected to be equal within about $1\%$,
since the variation of the two form factors from $q^2=m^2_{\pi}$ to 
$q^2=m^2_K$ is tiny for a relevant scale of order $m^2_B$.
We conclude that (once the smallness of rescattering has been established)
{\it the rates of these four processes can be used not only to determine the 
U-spin breaking factor in the ratio of amplitudes, but also to check the 
factorization assumption by finding equal ratios of amplitudes in the two 
cases.}

\section{Stringent constraints on weak phases}
In the present section we discuss recent developments in suggestions for 
determining the weak phase $\gamma$. We will also comment briefly on an old 
idea for resolving penguin uncertainties in the determination of $\sin 2\alpha$ 
from $B^0(t)\to \pi^+\pi^-$.
\subsection{$\phi_3=\gamma$ from $B^0,B_s\to K^{\pm}\pi^{\mp}$}
The processes in (\ref{SU3}) play a useful role in determining
$\gamma$. Here we describe a scheme based on $K\pi$ 
decays of $B^0$ and $B_s$ mesons \cite{bskpi}. We will briefly comment 
on a complementary method using the other two processes. 

Writing the amplitudes for $B^0\to K^+\pi^-$ and $B_s\to K^-\pi^+$ as in 
Eqs.~(\ref{s}) and (\ref{d}), respectively, we note that the rates for these
processes and their charge-conjugates depend on four quantities, 
$\vert V^*_{ub}V_{us}A_u\vert,~\vert V^*_{cb}V_{cs} A_c\vert,~\delta_{K\pi}
\equiv {\rm Arg}(A_u A^*_c)$ and $\gamma\equiv {\rm Arg}(-V^*_{ub}V_{ud}V_{cb}
V^*_{cd})$. Because of the equality of CP rate-differences in the two 
processes, a determination of $\gamma$ requires another input. This input
is provided by $|A(B^+\to K^0\pi^+)|=|V^*_{cb}V_{cs} A_c|$, where small 
rescattering corrections are neglected as argued above.

Defining two charge-averaged ratios of rates
\beq
R \equiv \frac{\Gamma(B^0 \to K^{\pm} \pi^{\mp})}
{\Gamma(B^\pm \to K \pi^{\pm})}~,~~~
R_s \equiv \frac{\Gamma(B_s \to K^{\pm} \pi^{\mp})}
{\Gamma(B^{\pm} \to K \pi^{\pm})}~~,
\eeq
and CP violating pseudo-asymmetries
\beq
{\cal A}_0 \equiv \frac{\Delta(B^0 \to K^+ \pi^-)}
{\Gamma(B^{\pm} \to K \pi^{\pm})}~,~~
{\cal A}_s \equiv \frac{\Delta(B_s \to K^- \pi^+)}
{\Gamma(B^{\pm} \to K \pi^{\pm})}~~,
\eeq
one finds
\beq \label{eqn:R}
R = 1 + r^2 + 2 r \cos \delta_{K\pi} \cos \gamma~~~,
\eeq
\beq \label{eqn:Rs}
R_s = \tl^2 + (r/\tl)^2 - 2 r \cos \delta_{K\pi} \cos \gamma~~~,
\eeq
\beq \label{A0s}
{\cal A}_0 = - {\cal A}_s = -2r \sin \delta_{K\pi}\sin \gamma~~,
\eeq
where $r\equiv \vert V^*_{ub}V_{us}A_u\vert/\vert V^*_{cb}V_{cs} A_c\vert$.
SU(3) breaking can be checked in (\ref{A0s}) and used for improving the 
precision in $\gamma$ obtained from these four quantities. 
It is estimated \cite{bskpi}
that a precision of $10^\circ$ in $\gamma$ can be achieved in experiments 
to be performed at the Fermilab Tevatron Run II program \cite{RUNII}. 

Alternatively, one may compare time-dependence in the U-spin 
related decays $B^0(t)\to \pi^+\pi^-$ and $B_s(t)\to K^+ K^-$ \cite{flei}. 
Here one is measuring in the two processes CP asymmetries of the form
\beq
Asym(t) = A_{\rm mix}\sin(\Delta mt) + A_{\rm dir}\cos(\Delta mt)~.
\eeq
The four measurables, $A_{\rm mix}$ and $A_{\rm dir}$ in the two processes,
can be expressed in terms of $\beta,~\gamma$, the ratio of penguin and tree
amplitudes in $B^0\to\pi^+\pi^-$ and their relative strong phase. This
allows a determination of $\gamma$ with a precision comparable 
to that achieved when studying $B,B_s\to K\pi$.

\subsection{$\phi_3=\gamma$ from $B^{\pm}\to K\pi$}
A large number of charmless $B$ and $B_s$ decays to two light pseudoscalars 
can be related to each other under approximate flavor SU(3) symmetry. It was 
noted a long time ago \cite{zepp} that hadronic weak amplitudes can be 
classified in SU(3) in terms of quark diagrams. Starting with the papers 
\cite{GHLR} this framework has been applied to the $\Delta B=1,~\Delta C=0$ 
low energy effective Hamiltonian (\ref{Heff}) and its $\Delta S=0$ counterpart
for the purpose of determining weak phases. 
A large number of proposals of this kind \cite{SU3} were made in the past 
seven years.

Here we will focus on a particular recent application.
Although SU(3) is only an approximate symmetry it will be applied 
to subleading terms in decay amplitudes so that SU(3) breaking corrections will 
be second order. We will make use of  
an SU(3) proportionality relation \cite{ewpVP} between electroweak penguin 
operators ($Q_{9.10}$) and the current-current operators ($Q_{1,2}$) in 
(\ref{Heff}) transforming as given SU(3) representations ($\overline{\bf 3},~
{\bf 6}$ and $\overline{\bf 15}$), in which the proportionality constant is 
given purely in terms of ratios of Wilson coefficients and CKM factors. For 
instance \cite{ewpVP}

\bea\label{15}
{\cal H}^{(s)}_{EWP}(\overline{\bf 15}) &=& -\frac32 \frac{c_9+c_{10}}{c_1+c_2} 
\frac{V^*_{tb}V_{ts}}{V^*_{ub}V_{us}} 
{\cal H}^{(s)}_{CC}(\overline{\bf 15})\nonumber\\
&=& -\delta_{EW}\ e^{-i\gamma}{\cal H}^{(s)}_{CC}(\overline{\bf 15})~,
\eea
where $(c_9+c_{10})/(c_1+c_2)  \approx -1.12\alpha,~\delta_{EW}=0.65\pm 0.15$. 
This SU(3) equality, implying a relation between hadronic amplitudes, 
simplifies the study of processes governed by $\overline{\bf 15}$ transitions. 

Where does only $\overline{\bf 15}$ contribute? The answer to this question
is simple \cite{NRPL}: In $B \to (K\pi)_{I=3/2}$ and in $B^+\to\pi^+\pi^0$ 
where the final states are ``exotic" and belong to a ${\bf 27}$ representation. 
The amplitude of the first proccess can 
be written in terms of SU(3) graphical contributions \cite{GHLR}
\bea\label{Kpi3/2}
& & A(B^+\to K^0\pi^+)+\s A(B^+\to K^+\pi^0)=\\
& & -(T + C + P_{EW} + P^c_{EW})=-(T + C)(1 - \delta_{EW}\ e^{-i\gamma})~.
\nonumber
\eea
Note that the $\Delta I =0$ penguin contributions dominating the 
two amplitudes in the left-hand-side are equal and cancel by isospin alone. 
Hence the resulting  SU(3) relation (\ref{Kpi3/2}) applies to the 
subdominant current-current and electroweak contributions.

One defines a charge-averaged ratio of rates  \cite{NRPL}
\beq\label{R*}
R^{-1}_*\equiv \frac{2[B(B^+\to K^+\pi^0) + B( B^-\to K^-\pi^0)]}
{B(B^+\to K^0\pi^+) + B(B^-\to \bar K^0\pi^-)}~.
\eeq
The amplitudes in the numerator and denominator involve a common dominant 
penguin amplitude and current-current and electroweak contributions 
which are related by (\ref{Kpi3/2}). Expanding in subdominant contributions
one derives the following inequality, to leading order in small quantities
\beq\label{bound}
|\cos\gamma - \delta_{EW}| \ge \frac{|1-R^{-1}_*|}{2\epsilon}~,
\eeq
where \cite{pipi}
\bea\label{eps}
& & \epsilon =\frac{|V^*_{ub}V_{us}|}{|V^*_{tb}V_{ts}|}\frac{|T+C|}{|P+EW|}
=\nonumber\\
& &\sqrt2 \frac{V_{us}}{V_{ud}}\frac{f_K}{f_\pi}
\frac{|A(B^+\to \pi^0\pi^+)|}{|A(B^+\to K^0\pi^+)|} = 0.20\pm 0.05~.
\eea
SU(3) breaking in subdominant terms is introduced through $f_K/f_{\pi}$.

A useful constraint on $\gamma$ follows for $R^{-1}_*\ne 1$. The error of
the present average value \cite{pipi}, $R^{-1}_*=1.45\pm 0.46$, ought to be 
reduced before drawing firm conclusions about allowed values of $\gamma$.
Further information about $\gamma$, applying also to the case $R^{-1}_{*}=1$, 
can be obtained by measuring separately $B^+$ and $B^-$ decay rates \cite{NRPRL}.
The solution obtained for $\gamma$ involves uncertainties due to SU(3) breaking
in subdominant amplitudes and an uncertainty in $|V_{ub}/V_{cb}|$, both of which 
affect the value of $\delta_{EW}$. Combined with errors in 
$\epsilon \propto |A(B^+\to\pi^+\pi^0)/A(B^+\to K^0\pi^+)|$, and in 
rescattering effects, one may expect to reach a precision in 
$\gamma$ as small as 10 or 20 degrees \cite{NRPRL}. 

\subsection{$\phi_3=\gamma$ from $B\to D K$}
In $B^+\to D K^+$ two amplitudes interfere due to color-favored $\bar b\to\bar 
c u \bar s$ and color-suppressed $\bar b\to \bar u c \bar s$ transitions. 
The relative weak phase between the two amplitudes is $\gamma$. 
We will describe two variants based in this useful property which permits
a measurement of $\gamma$~\cite{GW}. A brief discussion is included of recent 
progress made in studying relevant amplitudes and strong phases.

\medskip\noindent
~{\it (a) $B$ decay to $K$ and flavor specific $D^0$ modes}~\cite{ADS}

\medskip\noindent
The three-body decay $B^+ \to (K^-\pi^+)_D K^+$ involves an interference 
between two cascade amplitudes,
\beq
Aa_{K\pi} \equiv A(B^+\to D^0 K^+)A(D^0\to K^-\pi^+)~~,
\eeq 
and 
\beq
\bar A\bar a_{K\pi} \equiv A(B^+ \to \bar D^0 K^+)A(\bar D^0\to K^-\pi^+)~~.
\eeq
The first amplitude $A$ is color-suppressed and subsequently the 
$D^0$ decays into a Cabibbo-favored mode with amplitude $a_{K\pi}$. The second 
amplitude $\bar A$ is color-favored, 
and subsequently $\bar D^0$ decays with a doubly Cabibbo-suppressed (DCS) 
amplitude $\bar a_{K\pi}$. The
relative weak phase between $A$ and $\bar A$ is $\gamma$, their strong 
phase-difference will be denoted $\delta,~{\rm Arg}(A/\bar A)=\delta + 
\gamma$, and the relative phase between $a_{K\pi}$ and $\bar a_{K\pi}$ 
(including a relative weak phase $\pi$)
will be denoted $\Delta_{K\pi}\equiv {\rm Arg}(a_{K\pi}/\bar a_{K\pi})$. 
Omitting a common phase space factor,
\bea\label{KpiK}
\Gamma(B^+ & \to & (K^-\pi^+)_D K^+) = |Aa|^2 +|\bar A\bar a|^2 \nonumber\\
& + & 2|A\bar A a\bar a|\cos(\delta + \Delta + \gamma)~~,
\eea
where $a\equiv a_{K\pi},~\bar a\equiv \bar a_{K\pi},~\Delta\equiv 
\Delta_{K\pi}$.

The rate for the charge-conjugate process, $B^- \to (K^+\pi^-)_D K^-$, has a
similar expression in which $\gamma$ occurs with an opposite sign, while strong 
phases are invariant under charge-conjugation. The CP asymmetry in this process,
involving an interference of $Aa$ and $\bar A\bar a$, is 
proportional to $\sin(\delta + \Delta)\sin\gamma$, becoming maximal for
$|\bar A\bar a/Aa|=1,~\delta + \Delta=\pi/2$. 

Let us summarize the present updated information on the parameters appearing 
in Eqs.~(\ref{KpiK}). The DCS amplitude $\bar a$ was measured recently by CLEO
\cite{CLEODCS} and by FOCUS \cite{FOCUSDCS}, resulting in an average value
$|\bar a_{K\pi}/a_{K\pi}| = (1.23\pm 0.10)\tan^2 \theta_c = 0.063 \pm 0.005$. 
SU(3) symmetry predicts a value of $\tan^2\theta_c$ \cite{KTWZ} indicating 
some amount of SU(3) breaking in $\bar a/a$. Model-dependent studies of $\Delta$ 
suggest \cite{UD} that this phase, which vanishes in the SU(3) limit, can be 
as large as about $20^\circ$ or be even larger. Recently a method was suggested 
\cite{GGR} for measuring $\Delta$ at a charm factory. This phase plays an 
important role in 
studies of $D^0-\overline D^0$ mixing. Finally, the ratio $A/\bar A$ is 
estimated, $|A/\bar A| \sim 0.1$, using a CKM factor 
$|V^*_{ub}V_{cs}|/|V^*_{cb}V_{us}|\approx 0.4$ and a color-suppression
factor of about 0.25 measured in $B\to \bar D\pi$ decays \cite{BHP}. The latter 
measurements also indicate a small value for $\delta$.

We conclude that {\it the two amplitudes interfering in 
Eqs.~(\ref{KpiK}) are anticipated to be comparable in magnitude, 
$|\bar A\bar a/Aa|\sim 0.6$ and to involve a possibly large relative strong 
phase $\delta+\Delta$}. This is crucial for a feasible determination of 
$\gamma$ from the rate (\ref{KpiK}) and its charge-conjugate. To solve for
$\gamma$ requires observing another doubly Cabibbo-suppressed $D^0$ decay mode.
Such a study in the $K^+ \pi^- \pi^0$ channel is reported at this conference
\cite{smith}.
This method requires a large number of $B$'s, at least of order $10^8-10^9$
since ${\cal B}(B^+\to \bar D^0 K^+){\cal B}(\bar D^0\to K^-\pi^+) = (4.2 \pm 
1.4)\times 10^{-8}$.

\medskip\noindent
~{\it (b) $B$ decay to $K$ and $D^0$ CP-eigenstate modes}~\cite{GDK}

\medskip\noindent
Neglecting very small CP violation in $D^0-\bar D^0$ mixing, one can write 
neutral $D$ meson even/odd CP states (decaying, for instance, to 
$K^+K^-$ or $K_S\pi^0$) as $D^0_{\pm}=(D^0 \pm \bar D^0)/\s$. Consequently, 
one has up to an overall phase 
\beq
\sqrt{2}A(B^+\to D^0_{\pm} K^+) = \pm |\bar A| + |A|\exp[i(\delta + \gamma)]~~.
\eeq
Let us define charge-averaged ratios of rates for positive and negative CP 
states relative to rates corresponding to color-favored neutral $D$ flavor 
states
\beq\label{Rpm}
R_{\pm} \equiv \frac{2[\Gamma(B^+ \to D_{\pm} K^+) + \Gamma(B^- \to D_{\pm} 
K^-)]}{\Gamma(B^+ \to \bar D^0 K^+) + \Gamma(B^- \to D^0 K^-)}~~,
\eeq
and two corresponding pseudo-asymmetries
\beq\label{Apm}
{\cal A}_{\pm} \equiv \frac{\Gamma(B^+ \to D_{\pm} K^+) - \Gamma(B^- 
\to D_{\pm} K^-)}
{\Gamma(B^+ \to \bar D^0 K^+) + \Gamma(B^- \to D^0 K^-)}~~.
\eeq
These quantities do not require measuring the color-suppressed rate
$\Gamma(B^+ \to D^0 K^+)$ and its charge-conjugate. One finds
\bea\label{RA}
R_{\pm} &=& 1 + |A/\bar A|^2 \pm 2|A/\bar A|\cos\delta\cos\gamma~~,\nonumber \\
{\cal A}_- &=& -{\cal A}_+ = |A/\bar A| \sin\delta \sin\gamma~~.
\eea 

In principle, Eqs.~(\ref{RA}) provide sufficient information to determine the 
three parameters $|A/\bar A|, \delta$ and $\gamma$, up to certain discrete 
ambiguities. However, a value $|A/\bar A| \sim 0.1$ 
would be too small to be measured with good precision. 
One still obtains two interesting bounds 
\beq\label{LIMIT}
\sin^2\gamma \leq R_{\pm}~~,
\eeq
implying new constraints on $\gamma$.
Assuming, for instance, $|A/\bar A|=0.1,~\delta=0,~\gamma=40^\circ$, one finds 
$R_-=0.85$.
With $10^8~~B^+B^-$ pairs, using measured $B$ and $D$ decay branching
ratios, one estimates an error \cite{GDK} $R_-=0.85 \pm 0.05$. In 
this case, Eq.(\ref{LIMIT}) excludes the range $73^\circ 
<\gamma < 107^\circ$ with 90$\%$ confidence level. Including measurements of the
CP asymmetries ${\cal A}_{\pm}$ could further constrain $\gamma$. 

\subsection{$\phi_2=\alpha$ from $B\to\pi\pi$}
The phase $\alpha = \pi-\beta-\gamma$ occurs in the time-dependent rate of 
$B^0(t)\to \pi^+\pi^-$ and would dominate its asymmetry if only a ``tree" 
amplitude $T$ contributes. A smaller penguin amplitude $P$, 
which carries a different weak phase, implies a more general form of the 
time-dependent asymmetry, which includes in addition to the $\sin(\Delta mt)$ 
term a $\cos(\Delta mt)$ term due to direct CP violation  \cite{MG}
\beq\label{asymmet}
{\cal A}(t) = a_{\rm dir}\cos(\Delta mt) + \sqrt{1-a^2_{\rm dir}} 
\sin 2(\alpha + \theta)\sin(\Delta mt)~.
\eeq
This provides two equations for three unknowns, 
$a_{\rm dir},~\theta$ and $\alpha$, which is insufficient for measuring $\alpha$.
$a_{\rm dir}$ and $\theta$ can be expressed in terms of $|P/T|,~{\rm Arg}(P/T)$ 
and $\alpha$. Consequently, knowledge of $|P/T|$ or $\theta$ could provide very 
useful information about $\alpha$. Applying flavor SU(3) to
measured $B\to \pi\pi$ and $B\to K\pi$ decay rates \cite{pipi} one finds 
\cite{DGR} $|P/T| =0.3 \pm 0.1$. QCD based studies
of $|P/T|$ \cite{BBNS,AD} obtain a small value around 0.1, however
these calculations involve systematic theoretical uncertainties \cite{AK}.

A clean way of eliminating the penguin effect \cite{GRLO} is by measuring also 
the time-integrated rates of $B^0\to\pi^0\pi^0$, $B^+\to\pi^+\pi^0$
and their charge-conjugates. One constructs the isospin triangle
\beq\label{iso}
A(B^0\to\pi^+\pi^-)/\s + A(B^0\to\pi^0\pi^0) = A(B^+\to \pi^+\pi^0)
\eeq
and its charge-conjugate in which $A(B^+\to \pi^+\pi^0)$ is a common base.
The correction $2\theta$ in (\ref{asymmet}) is given by the angle between 
$A(B^0\to\pi^+\pi^-)$ and $A(\bar B^0\to\pi^+\pi^-)$. A tiny electroweak 
penguin term, forming a very small angle between $A(B^+\to \pi^+\pi^0)$ and 
$A(B^-\to \pi^-\pi^0)$, can be taken into account analytically \cite{GPY}.

A small $B\to\pi^0\pi^0$ branching ratio (probably around $10^{-6}$ but
hard to estimate) may be a potential difficulty for separating $B^0$ and
$\bar B^0$ decays into $\pi^0\pi^0$. A combined rate measurement
for $B^0$ and $\bar B^0$, avoiding the need for flavor tagging, is considerably
easier. Such a measurement was shown \cite{GQ} to imply useful upper bounds 
on $\theta$ if the combined rate is sufficiently small. An interesting question 
\cite{GLSS} is whether it can also lead to a lower bound on $\theta$, thereby 
constraining this angle tightly to a narrow range (and consequently fixing 
$\alpha$) for some values of the combined $\pi^0\pi^0$ rate and the other 
measurables.

\section{Conclusion}
In our brief conclusion we wish to make a few general recommendations for 
experimentalists. The experimental task becomes harder as we go down the list.
\begin{itemize}
\item Measure {\it crudely} as many as possible CP asymmetries in hadronic 
and radiative $B$ decays. Relative signs of U-spin related asymmetries, which 
are relatively easily measured, test the CKM picture. 
\item Certain asymmetries are predicted by CKM to be very small. 
Measuring sizable asymmetries in these channels would be signals of new physics.
\item Measure {\it precisely} certain charge-averaged rates, which imply 
interesting constraints on $\phi_3 = \gamma$.
\item Measure {\it precisely} those CP asymmetries from which $\phi_1=\beta,
\phi_2=\alpha$ and $\phi_3=\gamma$ can be determined in an accurate manner.
\end{itemize}
The final goal of measuring CP asymmetries in different processes is to 
test and overconstrain the CKM parameters in a manifold and critical manner, 
thereby opening a window into new physics.

\section*{Acknowledgments}
This work was supported in part by the Israel Science Foundation
founded by the Israel Academy of Sciences and Humanities,
and by the U. S. -- Israel Binational Science Foundation through Grant
No.\ 98-00237.

% Journal and other miscellaneous abbreviations for references
\def \ajp#1#2#3{Am. J. Phys. {\bf#1}, #2 (#3)}
\def \apny#1#2#3{Ann. Phys. (N.Y.) {\bf#1}, #2 (#3)}
\def \app#1#2#3{Acta Phys. Polonica {\bf#1}, #2 (#3)}
\def \arnps#1#2#3{Ann. Rev. Nucl. Part. Sci. {\bf#1}, #2 (#3)}
\def \art{and references therein}
\def \cmts#1#2#3{Comments on Nucl. Part. Phys. {\bf#1}, #2 (#3)}
\def \cn{Collaboration}
\def \ite{{\it et al.}}
\def \cp89{{\it CP Violation,} edited by C. Jarlskog (World Scientific,
Singapore, 1989)}
\def \epjc#1#2#3{Eur.~Phys.~J.~C {\bf #1}, #2 (#3)}
\def \epl#1#2#3{Europhys.~Lett.~{\bf #1}, #2 (#3)}
\def \ib{{\it ibid.}~}
\def \ibj#1#2#3{~{\bf#1}, #2 (#3)}
\def \ijmpa#1#2#3{Int. J. Mod. Phys. A {\bf#1}, #2 (#3)}
\def \jpb#1#2#3{J.~Phys.~B~{\bf#1}, #2 (#3)}
\def \jhep#1#2#3{JHEP {\bf#1}, #2 (#3)}
\def \mpla#1#2#3{Mod. Phys. Lett. A {\bf#1}, #2 (#3)}
\def \nc#1#2#3{Nuovo Cim. {\bf#1}, #2 (#3)}
\def \np#1#2#3{Nucl. Phys. {\bf#1}, #2 (#3)}
\def \pisma#1#2#3#4{Pis'ma Zh. Eksp. Teor. Fiz. {\bf#1}, #2 (#3) [JETP Lett.
{\bf#1}, #4 (#3)]}
\def \pl#1#2#3{Phys. Lett. {\bf#1}, #2 (#3)}
\def \pla#1#2#3{Phys. Lett. A {\bf#1}, #2 (#3)}
\def \plb#1#2#3{Phys. Lett. B {\bf#1}, #2 (#3)}
\def \pr#1#2#3{Phys. Rev. {\bf#1}, #2 (#3)}
\def \prc#1#2#3{Phys. Rev. C {\bf#1}, #2 (#3)}
\def \prd#1#2#3{Phys. Rev. D {\bf#1}, #2 (#3)}
\def \prl#1#2#3{Phys. Rev. Lett. {\bf#1}, #2 (#3)}
\def \prp#1#2#3{Phys. Rep. {\bf#1}, #2 (#3)}
\def \ptp#1#2#3{Prog. Theor. Phys. {\bf#1}, #2 (#3)}
\def \ptwaw{Plenary talk, XXVIII International Conference on High Energy
Physics, Warsaw, July 25--31, 1996}
\def \rmp#1#2#3{Rev. Mod. Phys. {\bf#1}, #2 (#3)}
\def \rp#1{~~~~~\ldots\ldots{\rm rp~}{#1}~~~~~}
\def \stone{{\it $B$ Decays} (Revised 2nd Edition), edited by S. Stone
(World Scientific, Singapore, 1994)}
\def \yaf#1#2#3#4{Yad. Fiz. {\bf#1}, #2 (#3) [Sov. J. Nucl. Phys. {\bf #1},
#4 (#3)]}
\def \zhetf#1#2#3#4#5#6{Zh. Eksp. Teor. Fiz. {\bf #1}, #2 (#3) [Sov. Phys. -
JETP {\bf #4}, #5 (#6)]}
\def \zpc#1#2#3{Zeit. Phys. C {\bf#1}, #2 (#3)}
\def \zpd#1#2#3{Zeit. Phys. D {\bf#1}, #2 (#3)}
\def \PDG{Particle Data Group, D. E. Groom \ite, \epjc{15}{1}{2000}}

\end{document}